# Tools for quantum simulation with ultracold atoms in optical lattices

*Florian Schäfer*[1*], *Takeshi Fukuhara*[2], *Seiji Sugawa*[3,4], *Yosuke Takasu*[1] *and Yoshiro Takahashi*[1]

Author addresses

[1]Department of Physics, Graduate School of Science, Kyoto University, Kyoto, Japan.
[2]RIKEN Center for Emergent Matter Science (CEMS), Saitama, Japan.
[3]Institute for Molecular Science, National Institutes of Natural Sciences, Okazaki, Japan.
[4]SOKENDAI (The Graduate University for Advanced Studies), Okazaki, Japan.
**e-mail: schaefer@scphys.kyoto-u.ac.jp*

Abstract | After many years of development of the basic tools, quantum simulation with ultracold atoms has now reached the level of maturity where it can be used to investigate complex quantum processes. Planning of new experiments and upgrading existing set-ups depends crucially on a broad overview of the available techniques, their specific advantages and limitations. This Technical Review aims to provide a comprehensive compendium of the state of the art. We discuss the basic principles, the available techniques and their current range of applications. Focusing on the simulation of varied phenomena in solid-state physics using optical lattice experiments, we review their basics, the necessary techniques and the accessible physical parameters. We outline how to control and use interactions with external potentials and between the atoms, and how to design new synthetic gauge fields and spin–orbit coupling. We discuss the latest progress in site-resolved techniques using quantum gas microscopes, and describe the unique features of quantum simulation experiments with two-electron atomic species.

**Key points**

- Quantum simulation with ultracold atomic gases in optical lattices can be used to study condensed-matter quantum many-body systems, which are hard to simulate with conventional computers.
- The control of interatomic interactions is key to successful quantum simulation, and it can be implemented at short range and long range through various methods.
- Non-equilibrium phenomena can be studied by using controlled dissipation or lattice perturbations.
- Quantum gas microscopes currently offer the most precise tool for the manipulation and readout of optical lattice quantum simulators.
- The use of artificial gauge fields enables the simulation of charged particle physics; furthermore, non-trivial effects are accessible through use of spin–orbit coupling, topological lattices and synthetic dimensions.
- Going from alkaline-earth-metal to two-electron alkaline-earth-metal-like atoms allows the study of SU($N$) symmetric systems.

## Introduction

Quantum simulation is an approach for studying quantum systems experimentally by using other controllable quantum many-body systems[1]. Ultracold atomic gases have become a well-established experimental platform for quantum simulation owing to the excellent controllability of the system parameters and refined measurement techniques[2,3]. Quantum simulation with ultracold atoms in optical lattices, in particular, benefits from a wealth of theoretical and experimental tools and can be applied to many fields, ranging from condensed matter physics and statistical mechanics to high-energy physics and astrophysics[4–6] (Fig. 1). The excellent tunability and controllability of the system parameters of ultracold atomic gases enables access to phenomena or regimes unavailable in other systems, such as the realization of the Bardeen–Cooper–Schrieffer to Bose–Einstein condensate (BCS–BEC) crossover or the generation of strong effective magnetic fields through artificial gauge fields[7,8].

In this Technical Review, we mainly focus on the application to solid-state physics whose models are naturally realized with ultracold atoms in optical lattices — even though at first glance the key parameters of both systems differ at times by more than ten orders of magnitude (Table 1) — and describe the tools used in these experiments. Real solid-state materials have many complex degrees of freedom, such as defects, impurities and multiple energy bands. In some cases, however, the essential features of the system are captured by a minimal theoretical model, an important example being the single-band Fermi–Hubbard model for high-$T_c$ cuprate superconductors. It is especially important to explore the underdoped region of the Fermi–Hubbard model where the origin of high-temperature cuprate superconductors could be discovered[9,10]. Numerical simulation methods are not powerful enough to simulate the Fermi–Hubbard model away from half filling[11,12]. In a quantum simulation approach, experiments using ultracold atoms in an optical lattice are performed to simulate the Fermi–Hubbard model itself, instead of the complex real solid-state materials.

This Technical Review provides an accessible source of technical references especially targeted at newcomers to the field of experimental quantum simulation with ultracold atoms. The article is structured into six main topics, each covering a particularly important main aspect of ultracold atom experiments towards quantum simulation (optical lattice basics and techniques, control of interatomic interactions, engineered perturbations, high-resolution imaging, synthetic gauge fields and spin–orbit coupling, and two-electron atoms). We break down each topic into the individual techniques, describe the methods involved and offer exemplary applications.

As we focus on quantum simulations using optical lattices, we could not include, or could only briefly mention, many other important topics, such as the BCS-BEC crossover; the physics of universal few-body bound states; experiments in box potentials; atom–ion hybrid systems; BECs of photons, polaritons or excitons; cavity-mediated interactions; the physics of lower-dimensional systems; quantum droplets and supersolids; quantum thermalization; quantum transport in narrow wires; or other developments, including spontaneous matter-wave emission.

## Optical lattices

An optical lattice — a periodic potential with the lattice spacing on the order of the laser wavelength — is a versatile tool to perform quantum simulations (Box 1). Analogous to the lattice structure of solid-state systems, an optical lattice imprints a well-defined structure onto the atomic cloud and serves as the reference frame to define inter-atomic interactions. The utility of such a system for the study of, for example, the superfluid-to-insulator phase transition was first recognized more than 20 years ago[13]. In this section, starting from the well-established procedure to prepare cold atoms in an optical lattice, we review how to emulate different systems (Hubbard, Heisenberg and Ising models) in optical lattices of various lattice geometries. We then discuss how the flexibility of cold atom systems allows us to perform 'protocols' — that is, sequences of combined system controls and measurements — to gain access to physical quantities otherwise difficult to obtain.

***Optical lattice basics.***

After formation of an ultracold atomic sample (see Supplementary Information Section S1 for a concise review of the process), the atoms are loaded into an optical lattice. Although periodic optical light fields can be created using various methods (discussed briefly below), we largely consider standing waves generated by counterpropagating laser beams, which is still the most important technique to the field. Depending on the laser wavelength, the atoms in an optical lattice are trapped in either the nodes or the antinodes by the optical dipole force. Such a periodic potential produced by an optical lattice gives rise to a series of Bloch bands (Fig. 2a). We note that before transferring the atoms into an optical lattice, it is possible to cool them in the harmonic trap to sufficiently low temperatures such that only the lowest Bloch band is naturally populated after adiabatic loading of the atomic sample into the optical lattice. When the lattice potential is sufficiently deep, the tight-binding model[14], in which an atom is localized at each lattice site and undergoes hopping between adjacent lattice sites, is applicable. In this situation, the interaction energy at a single site is much smaller than the energy gap between the ground state and the first excited band. The system can then be described by the Hubbard model, which includes on-site interactions, tunnelling and external confinement. At large enough on-site interactions, compared with the tunnelling energy with unity filling, the Hubbard model can be rewritten as spin Hamiltonians[15], such as the Heisenberg or Ising models. Spin–spin interactions in the Heisenberg model arise through super-exchange interactions. Dipole–dipole interactions (magnetic or electric) are caused by magnetic atoms or polar molecules[15,16]. Ising-type interactions are due to the mapping between spin and density in the Bose–Hubbard model[17,18] (Box 1).

Numerous many-body phases in solid-state systems arise from the competition between the various energy scales involved. The choice of the lattice geometry therefore has a crucial role in the design of a target quantum system. First, the lattice dimensionality (one[19], two[20] or three dimentions[4]) has a strong impact on the available many-body phases and their phase transitions. In low dimensions, quantum effects are generally enhanced by strong quantum fluctuations.; the 2D Fermi–Hubbard model is a prominent example. Second, each lattice configuration in real space leads to a unique energy band structure (Fig. 2). In the excited $P$-band of a square lattice, unconventional superfluidity is found[21]. In the Lieb lattice[22] (Fig. 2b), a dispersionless flat band appears, in which interactions significantly dominate over kinetic energy. In the honeycomb lattice[23] (Fig. 2c), which is analogous to graphene, Dirac cones appear in the energy band and topological physics can be explored. Further specialized lattice types, such as triangular[24] (Fig. 2d) or kagome-lattice systems[25] (Fig. 2e), can exhibit geometric frustration[26] in their ground states, which, due to strong quantum fluctuations, can be highly-entangled states. Moreover, by trapping multiple atomic species or states, species-selective potentials can be used to implement state-dependent[27,28] or mixed-dimensional lattices[29–32], in which, 'mediated interactions', for example, can be engineered for realizing unconventional pairings. Even more exotic lattices, such as quasi-crystals[33] and lattices within optical cavities[34] can also be realized to simulate unique physical systems. Finally, optical superlattices have many applications, from creating isolated double wells[35] to exploring topological physics[36,37].

The manipulation of the optical potential and the creation of optical lattices are not limited to standing waves of light. Holographic methods using masks or spatial light modulators[38,39], as well as diffractive optics using digital micromirror devices (DMDs) or acousto-optic deflectors , are also used to create and control optical potentials[40–42]. Furthermore, the above techniques can be used to form arrays of single atoms contained in microtraps created by tightly focused laser beams, so-called optical tweezers. By combining non-destructive and highly sensitive imaging methods with the targeted movement of selected tweezers, defect-free atomic arrays with spacings of only a few micrometres can be prepared in one, two and three dimensions[40,43,44].

In general, the external-light-field-induced polarizabilities and energy shifts (namely the light or ac Stark shift) of two different atomic states are not equal. Harnessing the light shift as a tool, it is possible to create spin-dependent lattices wherein the vector and tensor light shifts are dominant over the scalar light shift. Conversely, in some situations, it is possible to tune the trap or lattice lasers to a so-called magic wavelength at which the polarizabilities of both states become equal, and thus the difference in the light shifts vanishes. In this situation, it becomes feasible to investigate minute energy shifts, such as collisional shifts[45–49] and smaller perturbations.

***Controllable parameters.***

The fundamental parameters of the Hubbard model[4,50] (Box 1), namely the hopping matrix element (also often referred to as the hopping or tunnelling amplitude) and on-site interaction strength, can be precisely controlled experimentally. These parameters depend on the depth of the optical lattice potential, and their ratio, in particular, is finely controllable by changing the lattice depth. In addition to the ratio, the strength and sign of the on-site interactions can be controlled through Feshbach resonances (discussed further below). The hopping matrix elements can also be controlled by lattice shaking methods[51,52]. Although these matrix elements are usually real numbers, it is possible to induce complex hopping matrix elements, characterized by Peierls phases, using lattice shaking[53] and Raman-assisted tunnelling[54] methods (discussed below).

The filling factor (that is, the number of particles per lattice site) and temperature are also important parameters and are controllable by adjusting the total atom number and the initial entropy in a harmonic trap before adiabatically ramping up the lattice depth. As the laser beams forming the optical lattice usually have a Gaussian profile, a weak, overall harmonic trapping potential is superimposed on the lattice geometry. This additional potential generally leads to unavoidable inhomogeneities in the atom density. To overcome this issue, laser light tuned to create repulsive potentials can be used to create (quasi)uniform optical box traps[55]. Recent developments in advanced light-shaping techniques, such as DMDs and quantum gas microscopy techniques, also enable this limitation to be overcome for 1D and 2D gases.

***Methods to diagnose optical lattice systems.***

A rich set of tools for probing an optical lattice system is available. Of these, the time-of-flight (TOF) method is probably the most widely used. In the framework of optical lattice experiments, TOF images include information on the atomic coherence over the lattice sites[4]. Pioneering work revealed the superfluid-to-Mott-insulator quantum phase transition of the Bose–Hubbard model by observing the vanishing sharp interference peaks in TOF images[4,56]. These images show the 'real' momentum distribution of trapped atoms. However, the kinetic energy in periodic potentials is often discussed within the theory of Bloch bands in terms of Bloch wavefunctions and Brillouin zones, in which case, the quasi-momentum is then the relevant physical quantity. Quasi-momentum distributions of the atoms in multiple Bloch bands can be measured using the so-called band-mapping method after adiabatic ramp-down of the optical lattice followed by TOF imaging[20,57].

Various spectroscopic methods allow us to probe the band structures and properties of interacting and non-interacting atoms in an optical lattice. Band structures are often measured using two-photon $\Lambda$-type excitations, whereby two light beams with frequencies $f_1$ and $f_2$ with the associated wavenumbers $\mathbf{k}_1$ and $\mathbf{k}_2$, respectively, excite an atomic state with energy $E$ and quasi-momentum $\mathbf{k}$ to a state of energy $E \pm \Delta E$ and quasi-momentum $\mathbf{k} \pm \Delta\mathbf{k}$, where $\Delta E = h(f_1 - f_2)$ and $\Delta\mathbf{k} = \mathbf{k}_1 - \mathbf{k}_2$. Spectroscopy on a transition within the same band is often referred to as Bragg spectroscopy[58,59]. By contrast, lattice-modulation spectroscopy, which employs the temporal modulation of the lattice potential depth, can also excite the system between states with the same quasi-momentum, that is, $\Delta\mathbf{k} = 0$, and is often used to investigate higher Bloch band structures. This approach also allows the study of interactions, owing to their impact on the excitation spectrum[60]. As first demonstrated for an interacting ultracold Fermi gas in a trap without a lattice[61] and recently extended to the attractive Fermi–Hubbard model[62], angle-resolved photoemission spectroscopy (ARPES) can be used to probe the pairing of fermions and, in particular, the pseudo-gap, which is of great importance to the understanding of high-temperature superconductivity[62]. This ARPES method has been enabled by combining four basic steps: initial radio-frequency excitation of the interacting system to a non-interacting excited state, followed by band mapping of the quasi-momentum distribution of the excited atoms. A quantum gas microscope (discussed further below) is then used to measure the site-resolved atom distribution after conversion of atom momentum to position in real space using a harmonic trap[62]. In a related approach, the use of Raman spectroscopy has been proposed to obtain information on the Fermi surface of strongly correlated states[63].

The local density distribution is another useful physical quantity to diagnose optical lattice systems. The double occupancy in lattice sites is accessible by either observing the two-body loss after molecular creation[64] or by direct absorption imaging combined with high-resolution radio-frequency spectroscopy[65]. Multiple occupancies can also be revealed with high-resolution spectroscopy using radio-frequency[66] or optical clock transitions[45]. Recently, the internal energy of the Bose–Hubbard model was measured by combining TOF and site-occupancy measurements[46]. Last, but not least, the development of single-site imaging techniques, so-called quantum gas microscopes, has provided direct access to the in-situ atom distribution[39].

One of the advantages of a cold atom system is the flexibility of combining several controls and measurements, that is, it is possible to measure the system after having performed some local operations. In the following, we will refer to such sequences of operations and measurements as measurement protocols. Although many protocols have been proposed and demonstrated, we highlight here just a few key examples. By applying a spin-dependent potential gradient just before a TOF measurement, the spin components are separately imaged (magnetic[67] or optical[68] Stern–Gerlach imaging). For complex lattice geometries containing sublattices, such as a double-well or a Lieb lattice (Fig. 2b), the occupation numbers of each sublattice are also accessible by prior conversion into band populations[22,35]. The spin correlations between nearest neighbours at unity filling can be detected exploiting a singlet–triplet–oscillation protocol[69–71] (see Supplementary Information Section S2). Experiments that were based on the Talbot effect and combined in-trap atom expansion and thermalization after rapid optical lattice ramp-up succeeded in detecting non-local atom correlations and long-range coherences[72]. Measurement protocols to assess the Berry curvature and various related topological invariants have also been experimentally realized[23,73]. For example, in a recently proposed and demonstrated method, the excitation rate to higher Bloch bands by amplitude modulation of a position-dependent external potential, measured through a band-mapping technique, directly provided the real and imaginary parts of a quantum geometric tensor[74]. Operation sequences are also applied for quantum state manipulation. For example, the 'square root of swap' $\sqrt{\text{SWAP}}$ gate can be implemented by use of a spin-dependent optical lattice[75]. Finally, in combination with quantum gas microscopes and local operations, even more complex protocols become feasible, such as the measurement of the entanglement entropy.

### Controlled interatomic interactions

A non-interacting lattice system can be described by single-particle eigenstates and calculated without fundamental difficulties. However, it is the interactions between components of a quantum system that bring the quantum simulator to life. It is useful to distinguish between short-range and long-range interactions, such as the contact and the dipole–dipole interaction. There are also interactions intrinsic to the system under study and dynamically controlled ones, such as magnetic moments and Feshbach resonances. In the following section, we highlight and compare the most prominent techniques to create and control atomic interactions of relevance for quantum simulation applications.

***Isotropic and short-range interactions.***

We consider here the collision between two unbound atoms of an ultracold gas. If the energy of this unbound scattering state (called 'entrance channel' or 'open channel') approaches the energy of a bound molecular state ('closed channel'), a Feshbach resonance occurs and considerable mixing between the entrance and the closed channels is possible[76]. If the energy of the bound state is controlled in the experiment, the strength of this isotropic interaction itself becomes adjustable. Most importantly, differences in magnetic moments of the open and closed channels allow for magnetically tunable Feshbach resonances[76] (Fig. 3a). This is the workhorse method for precisely controlling interactions. However, the bound states are not limited to those in the ground electronic states, and it is possible to bridge the energy gap between the entrance channel and the bound state in the electronic excited state using laser light tuned near a photoassociation resonance, leading to optical Feshbach resonances[77–81] (Fig. 3b). Even for two-electron atoms (alkaline-earth-metal or alkaline-earth-metal-like atoms), for which fully occupied outer shells with vanishing total electronic spin seem to oppose magnetic tunability, subtle differences in the nuclear $g$-factor between the ground and excited states open the possibility of magnetically controlling interactions through orbital Feshbach resonances[82–84] in the case of extremely shallow binding energies, as for $^{173}$Yb.

Tight confinement of atoms in optical lattices leads to significant changes in the interaction dynamics of ultracold gases[85–87]. In a 1D system, there are transversally excited molecular bound states, and a confinement-induced resonance occurs when the 3D scattering length approaches the length scale of the transversal confinement[88–91] (Fig. 3c). This effect is not limited to single-species experiments, and has also been demonstrated with two-species experiments in mixed dimensions[30].

The four approaches to manipulate the short-range interactions (magnetic, orbital, optical and confinement-induced) discussed here can all be treated consistently in the Feshbach resonance framework. Magnetic control is most common and most readily achievable. In cases when magnetic control is not possible, other types of Feshbach resonances may offer a feasible approach. Optical control allows for extremely fast switching as well as submicrometre-scale control of the interactions, and confinement effects offer control of interactions under reduced dimensionalities.

***Anisotropic and long-range interactions.***

The resonance effects discussed above depend on the very close proximity of the scattering partners, leading to isotropic and short-range interactions. The inclusion of electromagnetic forces can lead to both long-range and anisotropic interaction effects. A prime example is the magnetic dipole–dipole interaction, which causes strong anisotropies in the interactions. These are particularly enhanced in atomic species with very large magnetic moments, such as Cr (ref.[92]), Dy (ref.[93]), Er (ref.[94]) and Ho (ref.[95]). Combined with the technique of Feshbach resonances, the relative strength of the isotropic, short-range interactions and the dipole–dipole interactions can be controlled[96]. In contrast to magnetic dipole moments, polar molecules comprising different atomic species[97,98] exhibit electric dipole moments, providing another approach towards anisotropic interactions. Two methods are pursued to generate cold polar molecules. Either the polar molecules are formed from laser-cooled cold atoms[99–102], or molecules are first created and then laser cooled to the required low temperatures[103–105]. By combining molecule association from ultracold atomic samples and further cooling techniques, it has even been possible to realize quantum-degenerate polar molecules[106].

Another important example is the electrostatic, long-range interaction provided by Rydberg atoms[107] that could pave the way to a Rydberg-based quantum computing infrastructure[108,109] and to several quantum simulation applications of the spin Hamiltonian[110], such as realizing the Ising model[111–115]. As building blocks, Rydberg blockade[116,117], Rydberg dressing of ground state atoms via off-resonant laser coupling[118] and dipole spin-exchange interactions[119] have been realized. Trapping of Rydberg atoms by the ponderomotive force in lattice potentials[120] and in blue-detuned hollow traps[121] allow for high fidelity control in future experiments using long-lived circular Rydberg atoms[122,123]. Finally, Rydberg states of two-electron atoms could offer further unique possibilities as they have an atomic structure that is different from that of alkali atoms[124].

## Controlled perturbations

In perturbing a quantum system, a changeover from a closed, equilibrated system to an open or non-equilibrium system is made possible. Controlled perturbations therefore broaden the range of accessible quantum simulation targets to go beyond equilibrium properties in the ground state. In this section, we focus on experimental methods to introduce perturbations through coupling to external degrees of freedom (dissipation) and through disordered potentials. Some possibilities offered by time periodic modulations are discussed in a later section. Additionally, we examine here how sudden changes of the system parameters (a quench) may be realized to create and study out-of-equilibrium situations.

***Dissipation.***

Although the previous section focused on elastic collisions, in the context of dissipation, inelastic collisions have an important role. The dissipation process can be classified on a microscopic level by the number of particles involved. One-body dissipation can be easily introduced by background-gas collisions in an uncontrollable way, whereas highly effective and controllable one-body dissipation is possible with near-resonant light, leading to heating by photon scattering events[125]. In a different approach, very localized dissipation, limited in its effects to just a single optical lattice site, has been achieved using tight electron beams[126]. When more than one particle is involved, the dissipation is governed by the collisional physics between atoms[127]. In this case, instead of driving transitions between two states of a single atom, dissipative coupling of two atoms in a photoassociation experiment to short-lived or untrapped molecular states is possible[128]. A controlled three-body dissipation has been demonstrated by tuning the scattering length to large negative values through a Feshbach resonance[129]. Inelastic processes are often detrimental to the long coherence times necessary for quantum simulation and computing applications. However, inelastic processes can also give rise to new effects that can be exploited as tools[130].

An example is the ‘a-watched-pot-never-boils’ quantum Zeno effect[131]. This effect has been experimentally studied in static optical lattice systems[132,128], wherein decay from a single state is suppressed. Moreover, in the absence of optical lattices, this effect has also been studied on whole subspaces[133,134], within which, in the theoretical framework of quantum Zeno dynamics[135], the wave function is free to evolve within only a part of the possible space of states. Theoretical studies have demonstrated that engineered dissipation can protect a system from decoherence caused by otherwise uncontrollable dissipative effects[136].

Dissipation control therefore makes it possible to switch from exploring the standard Hubbard model (Box 1) to dissipative lattice systems for both bosons and fermions. Two-body dissipation, for example, has been used in experiments as a tool to suppress the growth of phase coherence and to stabilize the Mott-insulator state in a dissipative Bose–Hubbard model[128] and in a dissipative Fermi–Hubbard model[137] in which a highly entangled Dicke state was created. Such dissipative Hubbard models are also predicted to lead to a dynamical change of the spin correlation[138]. Anomalous, subdiffusive momentum broadening due to dissipation has also been observed[139]. Considering a weak dissipative perturbation, a non-Hermitian version of the linear-response relation has recently been proposed[140]. A spatially dependent dissipation with a $\pi/2$ phase difference to the optical lattice potential can realize an interesting type of non-Hermitian Hamiltonian with parity and time-reversal symmetry, which is predicted to exhibit novel behaviour[141].

***Disorder potentials.***

In contrast to the predominantly temporal perturbation caused by dissipation, spatial perturbation owing to non-periodic potential landscapes enables the quantum simulation of disordered matter. A well-established route to disordered potentials is through optical potentials (Fig. 4). In its most basic sense, speckle patterns focused down to the very small length scales of optical lattice experiments provide access to the disordered regime[142] (Fig. 4a). Quasi-periodic optical lattices (Fig. 4b), superpositions of optical lattices at incommensurable lattice spacings, allow a degree of control of the disorder to be regained and have proved to be equally effective[143]. In a different approach, disorder is introduced by adding a minority population acting as impurities to the majority species (Fig. 4c). Beyond changes in the local energy landscape, inter-species atom–atom collisions have demonstrated the impact of small impurity admixtures on fundamental phenomena, such as the superfluid-to-Mott-insulator transition[144]. All three approaches — random speckles, quasi-periodic potentials and atomic impurities — have been instrumental to studies of Anderson localization phenomena[142,143,145] (Fig. 4d). In the presence of interatomic interactions and dissipation, many-body localized states can form that, although still far from equilibrium, cannot thermalize and thus remain insulating, even at nonzero temperature[146,147].

***Out-of-equilibrium dynamics.***

The accessible physics is broadened beyond steady-state properties by time-dependent changes of the system Hamiltonian. If these changes or perturbations of the system are performed very quickly with respect to the other relevant timescales, it is referred to as a quench. These quenches drive the atomic system out of equilibrium and provide access to the physics of the time dynamics in ultracold atom systems[148]. In an optical lattice set-up, for example, the lattice depth can be changed either nearly instantaneously or by a continuous, but still fast, sweep across a phase transition. In the latter case, the speed of variation is an additional parameter of the experiment. In both scenarios, non-equilibrium dynamics can be studied.

Changing the lattice depth from the deep Mott-insulator regime (Box 1) to the shallow superfluid gives access to the phase coherence dynamics of the system[149]. Thus, in extension to the Kibble–Zurek mechanism of quenches across classical phase trans-itions[150], the formation of excitations after entering the superfluid state[151] and the build-up of the coherence lengths[152] can be assessed. Limits on the propagation speed of correlation information, which are important for understanding a quantum many-body system, have also been obtained in quenched lattice experiments with a quantum gas microscope[153]. Other experiments demonstrated inhibited ballistic expansion of bosons in systems with reduced integrability[154]. Moreover, for fermionic quantum gases, the out-of-equilibrium dynamics after suddenly turning off a weak initial harmonic confinement has been investigated. A transition from ballistic expansion for a non-interacting quantum gas to diffusive expansion for an interacting system has been observed[155]. Studies of the mass transport in a two-component, 1D Fermi gas after a sudden release from an optical lattice with a harmonic trap potential along the 1D direction into a homogeneous lattice revealed phase separation between fast singlons and slow doublons[156,157]. This so-called quantum distillation could serve to dynamically create low-entropy regions in a lattice. In a different approach, using fast magnetic-field control and suitable Feshbach resonances, quenches of the scattering length are possible. In one such experiment, fast density fluctuations in a 2D BEC of Cs atoms was observed[158].

## Quantum gas microscope

An important feature of ultracold atom experiments is the capability of very precise manipulation and high-sensitivity detection. Quantum gas microscopes[39,159] enable us to observe and control atoms in optical lattices with single-atom sensitivity and single-site resolution. In this section, we describe several tools necessary for quantum gas microscopy experiments and examples of their applications.

### *Key technologies of quantum gas microscopes.*

In Hubbard-regime optical lattice systems, the lattice periods need to be short to obtain a sufficiently large hopping matrix element between adjacent lattice sites. This poses formidable challenges to the experimental set-up and the required imaging optics (see Supplementary Information Section S3). In addition to the choice of imaging optics, the imaging method also requires careful consideration. Available imaging techniques include standard absorption imaging, fluorescence imaging (Fig. 5a) and Faraday imaging. The latter two, in particular, merit closer examination. Although high-resolution absorption imaging has been developed for the detection of local properties, such as density or incompressibility[160], single-atom sensitivity and single-site resolution are still difficult to achieve owing to limited scattering cross sections and heating by photon scattering. Fluorescence imaging, as a highly sensitive and background-free method, could make it possible to obtain a sufficiently strong signal from a single atom. However, fluorescence imaging inevitably leads to destruction of the quantum state by recoil heating. In that respect, off-resonant Faraday imaging offers the possibility of minimally destructive detection, albeit squeezed light might be necessary to overcome limitations in the signal-to-noise ratio[161].

In order to deal with heating due to photon scattering, several cooling schemes are used (Fig. 5b–e). Molasses cooling is a standard technique for Rb atoms, which have well-separated hyperfine states in the excited state[39,159]. For Li and K atoms, electromagnetically induced transparency cooling[162,163] or Raman sideband cooling techniques have been used[164–166]. A combination of molasses cooling and sideband cooling on a narrow-line transition has been demonstrated for Yb atoms[167]. Even without cooling, an extremely deep optical lattice potential for the excited electronic state of the optical probing transition makes it possible to obtain a sufficiently large number of fluorescence photons before the sample is heated up[168](Fig. 5f).

Drawbacks of early versions of quantum gas microscopes are parity projection and insensitivity to spin components (that is, hyperfine states): If there is more than one atom per site, then pairwise atom loss occurs owing to light-assisted collisions caused by the near-resonant imaging light. Therefore, the observed quantity is the parity of the atom occupation in each lattice site and not the exact on-site atom number. The cooling beam also mixes up the hyperfine states during the many absorption cycles necessary for a sufficiently large fluorescence signal. A straightforward solution to achieve spin-selective detection is to apply a spin-selective removal procedure before fluorescence imaging. By spatially separating different spin components or atoms before fluorescence imaging, more advanced spin-resolved or atom-number-sensitive measurements have been realized[169–171].

By illuminating the sample through a high-resolution imaging system with a Gaussian-shaped laser beam, control of a quantum gas on the single-atom and single-site level has been successfully demonstrated[172]. More complex patterns of light, and therefore nearly arbitrary potentials, can be projected onto the atoms with the help of a spatial light modulator, such as a DMD[41,42]. Operation of a spatial light modulator in the Fourier plane allows one to correct aberrations in high-resolution imaging systems and thus to obtain ultra-precise light patterns[173], while the direct imaging configuration offers advantages in terms of experimental and numerical simplicity.

***Applications.***

Quantum gas microscopes provide a snapshot of a quantum many-body system, from which it is possible to extract correlations between atoms as well as their spatial distribution. Direct probing of the Mott-insulating state has been demonstrated for both bosonic[41,159] and fermionic[174,175] atoms. Particle–hole pairs, which stem from quantum fluctuations, have been measured in the Mott-insulating state with finite tunnelling[176]. Anti-ferromagnetic correlations[171,177,178] and ordering[179] have been observed in Fermi–Hubbard systems, a milestone in acquiring new insight into high-$T_c$ cuprate superconductors. String orders (that is, non-local correlations) are accessible[176] and have been used to reveal hidden anti-ferromagnetic correlations[180]. Quantum entanglement, which lies at the heart of quantum information processing, also characterizes the quantum phases and dynamics of many-body systems[181]. Growth and propagation of entanglement has been measured in a 1D spin chain[169]. Entanglement entropy has been probed using the interference of two copies of a many-body state[182]. Single-site and single-atom addressing techniques can be used to prepare specific initial states for investigating quantum walks of atoms[172,183] or spin wave propagations[42,184]. Non-equilibrium dynamics in isolated quantum systems is among the most fundamental problems in statistical physics, and in this direction, several intriguing phenomena have been observed, such as quantum thermalization[185] and many-body localization[186]. Cooling in optical lattices is a central issue in quantum simulations with optical lattice systems. Entropy redistribution, which is one possible candidate to overcome the issue, has been demonstrated by locally manipulating the optical potential[179,187].

## Synthetic gauge fields

Many fascinating phenomena in solids[188] that arise from the interaction of electrons with electromagnetic fields and spin–orbit coupling cannot be simulated directly owing to the charge neutrality of atoms. However, in the past decade, several advances have been made to artificially engineer such effects. In the following, we introduce tools to implement artificial gauge fields, spin–orbit coupling and topologically non-trivial bands[189] in optical lattices.

***Artificial magnetic fields and topological lattices.***

The basic idea to emulate a charged particle in a vector potential field leads back to the Aharonov–Bohm effect. When a charged particle moves around a solenoid, the particle acquires a phase proportional to the magnetic flux that penetrates the closed trace. This effect can be mimicked by the geometric phase acquisition of a quantum state. When atomic internal states are coupled with Raman lasers, laser-dressed atoms in a non-uniform magnetic field can acquire Berry phases owing to the underlying Berry gauge field[190]. This technique, which was originally realized in BECs, has also been applied to observe a Peierls phase in a lattice potential[54]. Alternatively, a Raman-assisted tunnelling technique that couples neighbouring lattice sites by resonant Raman transitions is also accompanied by a Peierls phase acquisition. This method does not require atomic internal degrees of freedom but needs site-dependent energy offsets, created either by magnetic field gradients or by superlattice potentials to suppress the bare tunnelling and to resolve the tunnelling resonance. The atoms moving around a plaquette, the smallest closed loop for the atoms in the lattice, can acquire a non-zero tunnelling phase that mimics the Aharonov–Bohm phase acquired around a plaquette with non-zero magnetic flux. This technique was used to realize staggered[191] and strong uniform magnetic fields[192–194]. The latter was then used to realize the topological Hofstadter model with a non-zero Chern number, measured using the anomalous-Hall-response-induced centre-of-mass motion[73]. An artificial gauge field can be engineered not only with a Raman laser, but also by periodically modulating the phase of the lattice potential off-resonantly with respect to the bandgap or the on-site interaction energies. This rapid shaking of the lattice induces an inertial force on the atoms with respect to the lattice frame. In the framework of Floquet theory, the fast modulation is averaged out and a Floquet–Bloch band describes the system in which a complex tunnelling matrix element is engineered. With this Floquet engineering, artificial gauge fields[53], staggered magnetic fields[26] and the topological Haldane model[23] have been realized. The lattice shaking approach has the advantage that it does not require an additional laser[194]. More recently, the technique was extended to engineer density-dependent gauge fields[195–197], a step towards the simulation of dynamical gauge fields, and to measure the Chern numbers in the Haldane model following a quench[198]. A detailed comparison of various synthetic gauge-field implementations is given in the Supplementary Information Section S4.

***Synthetic dimensions.***

The available dimensions are not limited to the spatial ones, but can also be represented by time, internal states or momentum space. The dynamical version of the quantum Hall effect, also known as the Thouless charge pump, is realized using time as the second dimension. Quantized centre-of-mass transport per cycle is observed for both bosonic and fermionic systems[36,37]. Using the internal degrees of freedom of atoms (for example, the Zeeman states in alkali-metal atoms) as artificial lattice sites, a synthetic dimensional lattice is realized, within which a chiral edge current has been observed[199,200]. The coupling of internal states by a laser beam mimics the tunnelling between neighbouring sites. Here, the Peierls phase along the artificial site direction depends on the real lattice site position through the position dependence of the phase difference between the optical lattice and the coupling laser. Thus, atoms that move around a plaquette of the synthetic 2D lattice can acquire a phase. A momentum-space lattice can be realized through the coupling of discrete momentum states. Using laser-coupled internal states as the second dimension, a 2D lattice with non-zero flux was engineered[201]. Different from a normal lattice, these synthetic-dimension approaches realize hard-wall boundary conditions with a limited number of sites along the artificial lattice direction, and the interactions along the artificial dimensions are non-local[202], establishing them as unique systems in the quantum simulation toolbox.

***Spin–orbit coupling.***

Spin–orbit coupling can be engineered through the Raman-coupling of internal states or by Raman-laser-assisted tunnelling in an optical lattice owing to spin–momentum locking[203–206]. Furthermore, optical Raman lattices[207–209] have been implemented to realize 2D spin–orbit coupling with topological bands and a 3D semi-metal. In the Raman-lattice scheme, two pairs of lasers simultaneously form the conventional lattice and the necessary Raman potentials to realize 2D spin–orbit coupling[210,211] (see Supplementary Information Section S4 for a comparison of different implementation schemes).

## Two-electron atoms

Compared with alkali-metal atoms, which have a single valence electron that governs the physics of interest, two-electron systems, such as alkaline-earth-metal and alkaline-earth-metal-like atoms (such as Yb) provide additional unique features. Among these, access to SU($N$) symmetry and two-orbital systems offer intriguing quantum simulation tools and techniques that are otherwise impossible to perform. In this section, we address the preparation and detection methods of such SU($N$) and two-orbital physics.

***SU(N) systems.***

The ground electronic state of two-electron atoms is represented by a term $^1S_0$, where both the electron spin and orbital angular momenta are zero. Although bosonic isotopes have no nuclear spin, fermionic isotopes of, for example, $^{87}$Sr, $^{171}$Yb and $^{173}$Yb have nonzero nuclear spins, $I$, of 9/2, 1/2 and 5/2, respectively. The fact that the spin degree of freedom in the $^1S_0$ state of fermionic two-electron atoms is solely attributable to the nuclear spins and that the interatomic potential scarcely depends on the nuclear spins results in a nearly ideal SU($N$) symmetry[212], where $N = 2I+1$. The unique quantum magnetic phases for a Fermi–Hubbard model with SU($N$) symmetry are extensively studied theoretically and expected to yield rich physics[212]. One straightforward consequence of the SU($N$) symmetry is the absence of spin-exchange collisions, which differs from the case of alkali-metal atoms and results in stable populations of each spin component[68,213]. This stability is advantageous in the implementation of synthetic dimensions using this large spin system[200].

The enlarged spin symmetry of SU($N$) can be a powerful tool to lower the temperature of atoms in an optical lattice, and is known as the Pomeranchuk cooling effect[214]. During the adiabatic loading of the atoms into the optical lattice, the total entropy is constant. At unity filling, for example, each localized atom can carry a large entropy in the spin degrees of freedom, resulting in cooling of the system. This Pomeranchuk cooling effect has been confirmed by doublon production-rate measurements[215], in situ density distributions[216] in a spin-uncorrelated Mott region at high temperatures, and antiferromagnetic spin-correlation measurements at low temperatures[70]. Special care needs to be taken when manipulating the nuclear spin degrees of freedom of the SU($N$) fermions, which are nearly 1,000 times less sensitive to external magnetic fields than electron spins. Instead of an external magnetic field widely used for alkali atoms, one can use a pseudo-magnetic field that originates from a spin-dependent light shift generated by an off-resonant circularly or linearly polarized light field[68,217]. Such a pseudo-magnetic field gradient has been used to measure spin populations in optical Stern–Gerlach measurements[68] and to optically induce nuclear spin singlet–triplet oscillations[70]. Note that this reduced sensitivity to external magnetic fields, combined with the availability of optical manipulation methods, is advantageous for quantum information processing applications.

***Two-orbital systems.***

The existence of long-lived metastable $^3P_0$ and $^3P_2$ electronic states in two-electron atoms gives rise to unique manipulation tools. The resulting ultranarrow optical transitions between the $^1S_0$ ground state and these metastable states (‘clock transitions’) can be a versatile tool for an occupancy-resolved spectroscopy[45–49] in which the on-site collisional shift is much larger than the spectral linewidth.

Furthermore, the existence of electronic angular momentum in the $^3P_2$ state provides a tool for tuning the interatomic interaction between atoms in the $^1S_0$ and $^3P_2$ states through a magnetic Feshbach resonance induced not only by isotropic interactions, but also by anisotropic interatomic interactions[218,219]. In the vicinity of a Feshbach resonance, the bound state is mixed with the scattering state[220], which can enhance the strengths of optical Feshbach resonances[221]. This enhancement of optical Feshbach resonances will become a versatile asset for controlling the ground-state interatomic interactions of two-electron atoms. Note that optical Feshbach resonances have already been demonstrated for the related $^1S_0$–$^3P_1$ transition[77,78], and because of the relatively narrow linewidth, efficient control with only small losses was realized.

The absence of electronic angular momentum in the $^3P_0$ state provides an experimental platform for a two-orbital $^1S_0 + {}^3P_0$ SU($N$) system, whose rich quantum phases were theoretically studied[222]. Interestingly, the orbital degrees of freedom and interorbital nuclear-spin-exchange coupling provide an SU($N$) symmetric orbital Feshbach resonance[82–84], similar to the magnetic Feshbach resonance of alkali atoms with electron spin degrees of freedom and hyperfine coupling. The observed molecular bound state in the $^1S_0 + {}^3P_0$ state[223] can be similarly exploited for use in optical Feshbach resonances.

The two-orbital $^1S_0 + {}^3P_0$ system is also proposed[224] as an ideal experimental base for studying spin-orbital physics, such as the Kondo effect[222,225], for which experimental efforts using $^{173}$Yb and $^{171}$Yb have been recently reported[226,227].

## Outlook

Ultracold atoms in optical lattices realize several theoretical models, such as the Hubbard, Heisenberg and Ising models, which are crucial in condensed matter physics. We have described various tools for the quantum simulation of these theoretical models and several applications for the quantum simulation of both numerically hard and/or conceptually important problems[228].

Finally, we briefly outline future directions as well as challenges and opportunities for quantum simulation with ultracold atoms in optical lattices. Although the currently achieved temperature is sufficiently low to study new behaviours, such as pseudogap phenomena of the Fermi–Hubbard model, one important technical issue is reaching low enough temperatures for fermionic atoms in an optical lattice, to enable the investigation of the underdoped region of high-$T_c$ cuprate superconductors. Note that the temperature of fermions in an optical lattice is on the order of nano-kelvin, which is much colder than the sub-kelvin temperatures for electrons in solids. However, the relevant quantity in this case is the temperature scaled by the hopping matrix element, $T/t$, which is on the order of 0.1 for ultracold atoms, whereas it is typically below $10^{-4}$ for electrons in solids (Table 1). Several schemes have been discussed[229–231]. Different lattice configurations predicting higher $T_c$ (Ref.[232]) would be interesting new targets for realizing unconventional high-$T_c$ superfluids. A further important direction is quantum computing using ultracold atoms, which can be pursued with a Rydberg atom tweezer array[233]. Noisy intermediate-scale quantum devices[234] would be an important near-term target. Many sophisticated tools, developed for quantum simulation, can also be applied to other fields, such as precision measurements with a Fermi degenerate optical lattice clock[47] . Future works will pursue other fundamental physics research[235], such as the search for new particles[235] using cold atoms or molecules in an collisional-energy-shift suppressing optical lattice.

## Acknowledgements

This work was supported through the Grants-in-Aid for Scientific Research (KAKENHI) program of the Ministry of Education, Culture, Sports, Science, and Technology (MEXT) and the Japan Society for the Promotion of Science (JSPS) (grant nos 25220711, 17H06138, 18H05405, 18H05228, and 19H01854), the Impulsing Paradigm Change through Disruptive Technologies (ImPACT) program, the CREST program of the Japan Science and Technology (JST) Agency (grant no. JPMJCR1673), and the MEXT Quantum Leap Flagship Program (MEXT Q-LEAP) (grant no. JPMXS0118069021). S.S. acknowledges support from the JST, PRESTO (grant no. JPMJPR1664) and JSPS (19K14639).

**Author contributions**

All authors have read, discussed and contributed to the writing of the manuscript.

**Competing interests**

The authors declare no competing interests.

Table 1 | **Comparison between solid-state and optical-lattice systems.** Typical values of the key physical parameters are described for electrons in solid-state systems and fermionic atoms in optical lattices.

| **Parameter** | **Electrons in solids** | **Fermionic atoms** |
|---|---|---|
| Spin | 1/2 | 1/2, 3/2, … |
| Mass | $\sim 10^{-30}$ kg | $10^{-26}$–$10^{-25}$ kg |
| Lattice constant | ~0.5 nm | ~500 nm |
| Tunneling rate / energy | $\sim 10^{14}$ Hz / $\sim 10^{4}$ K | 100–1,000 Hz / 5–50 nK |
| Interactions | Coulomb | Van der Waals, on-site |
| Density | $\sim 10^{23}$ cm$^{-3}$ | $10^{13}$–$10^{14}$ cm$^{-3}$ |
| Fermi temperature ($T_F$) | $\sim 10^{4}$ K | ~100 nK |
| Temperature | ~ 1 K ($\sim 10^{-4}\,T_F$) | ~10 nK ($\sim 0.1\,T_F$) |

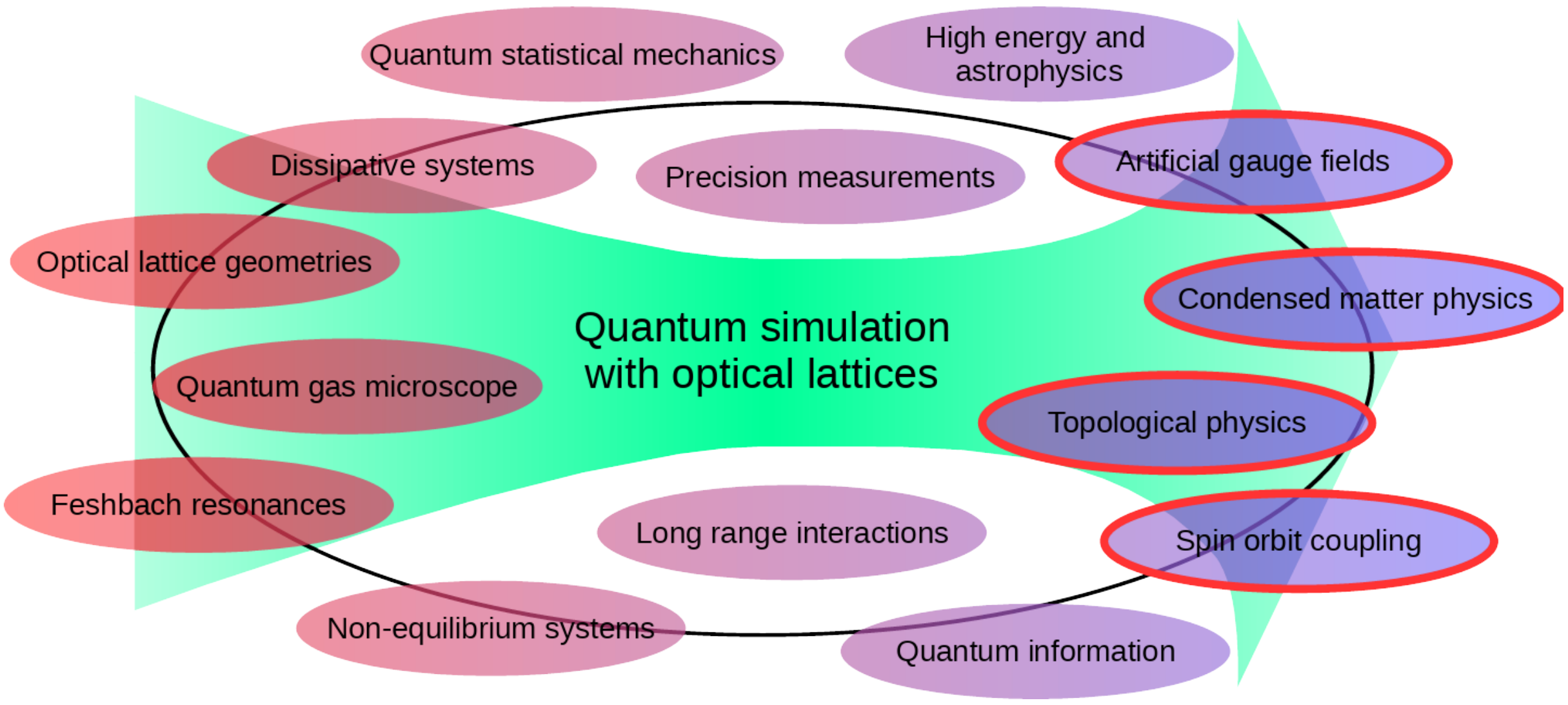

Fig. 1 | **Quantum simulation tools and applications.** Quantum simulation with optical lattices encompasses diverse fields that serve as tools, target applications or both. A clear distinction is often neither possible nor desirable. We give here an overview of the general fields and how, although all interconnected, they can be seen as tools to (red) and applications of (blue) quantum simulation approaches, with many topics positioned in-between these classifications. This Technical Review focuses on applications towards condensed matter physics and related fields (outlined in red).

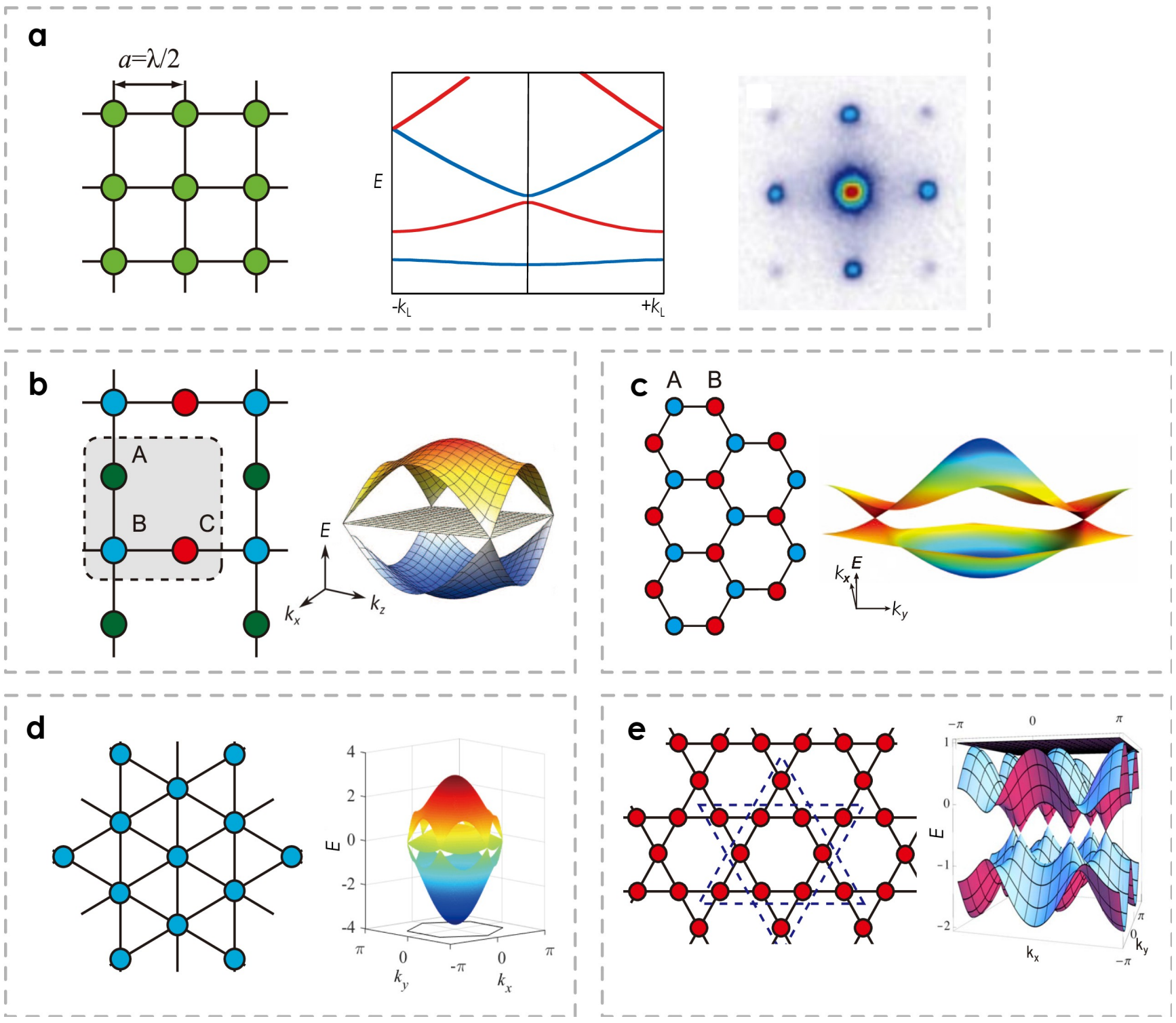

Fig. 2 | **Optical lattice geometries and Bloch band structures. a** | Regular square lattice configuration in real space (left) with the lattice spacing, *a*, being half the lattice laser wavelength, *λ*. The periodic potential leads to Bloch bands[85] (middle), where $k_L = \pi/a$ is the laser wavenumber.. The matter-wave interference patterns formed by a Bose–Einstein condensate after free expansion from a 3D cubic lattice reflects its momentum distribution[4] (right). **b** | Real-space lattice structure (left) and energy bands (right) of a Lieb optical lattice, for which a flat band appears in the first excited level[22]. A, B and C denote the three sublattices. **c** | Real-space honeycomb lattice (left). Dirac points appear in the band structure (right). A and B denote the two sublattices[236]. **d** | Triangular lattice in real space (left) and band structure with Dirac points (right)[237]. **e** | Real-space kagome lattice configuration (left) and band structure (right) with emerging Dirac cones and a flat band[238]. *E*, band energy; *k*, wavenumber of the wave packets. Panel **a** (centre) adapted from ref.[85], Springer Nature Limited. Panel **a** (right) adapted from ref.[4], Springer Nature Limited. Panel **b** (right) adapted with permission from ref.[22], AAAS. Panel **c** (right) adapted from ref.[236], Springer Nature Limited. Panel **d** (right) adapted with permission from ref.[237], APS. Panel **e** (right) adapted with permission from ref.[238], APS.

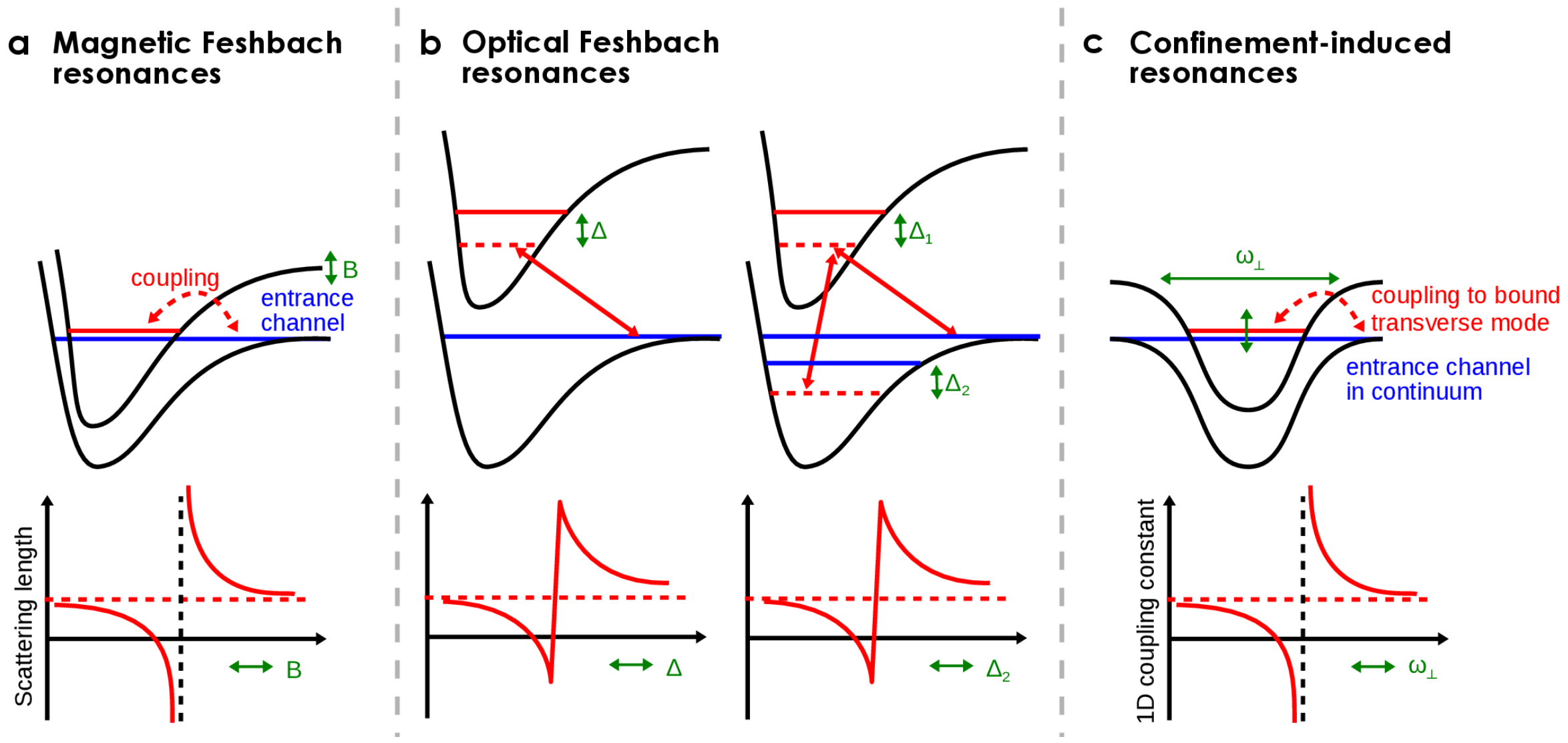

Fig. 3 | **Controlling atomic interactions using magnetic, optical and confinement-induced Feshbach resonances.** **a** | In a magnetic Feshbach resonance (top), the energy of a molecular bound state (red) is magnetically tuned to approach the low energy of the entrance channel (blue). The Feshbach resonance can modify the scattering length (bottom) over many orders of magnitude. **b** | In an optical Feshbach resonance (top), a similar modification of the scattering length (bottom) is achieved by bridging the energy gap (Δ) between the entrance channel and the bound state with suitably tuned laser light (red arrows). This can be achieved by using either a one-photon excitation scheme (left) or by driving a two-photon Raman transition[81] (right). **c** | A confinement-induced resonance of the 1D coupling constant occurs in a harmonic confinement (transverse oscillator frequency $\omega_\perp$) when the strength of confinement is tuned (horizontal green arrows) such that the energy of the incident channel in the continuum matches (vertical green arrow) the energy of a transversally excited bound state[90]. Panels **a** and **b** adapted courtesy of Johannes Hecker Denschlag, University of Ulm, Germany.

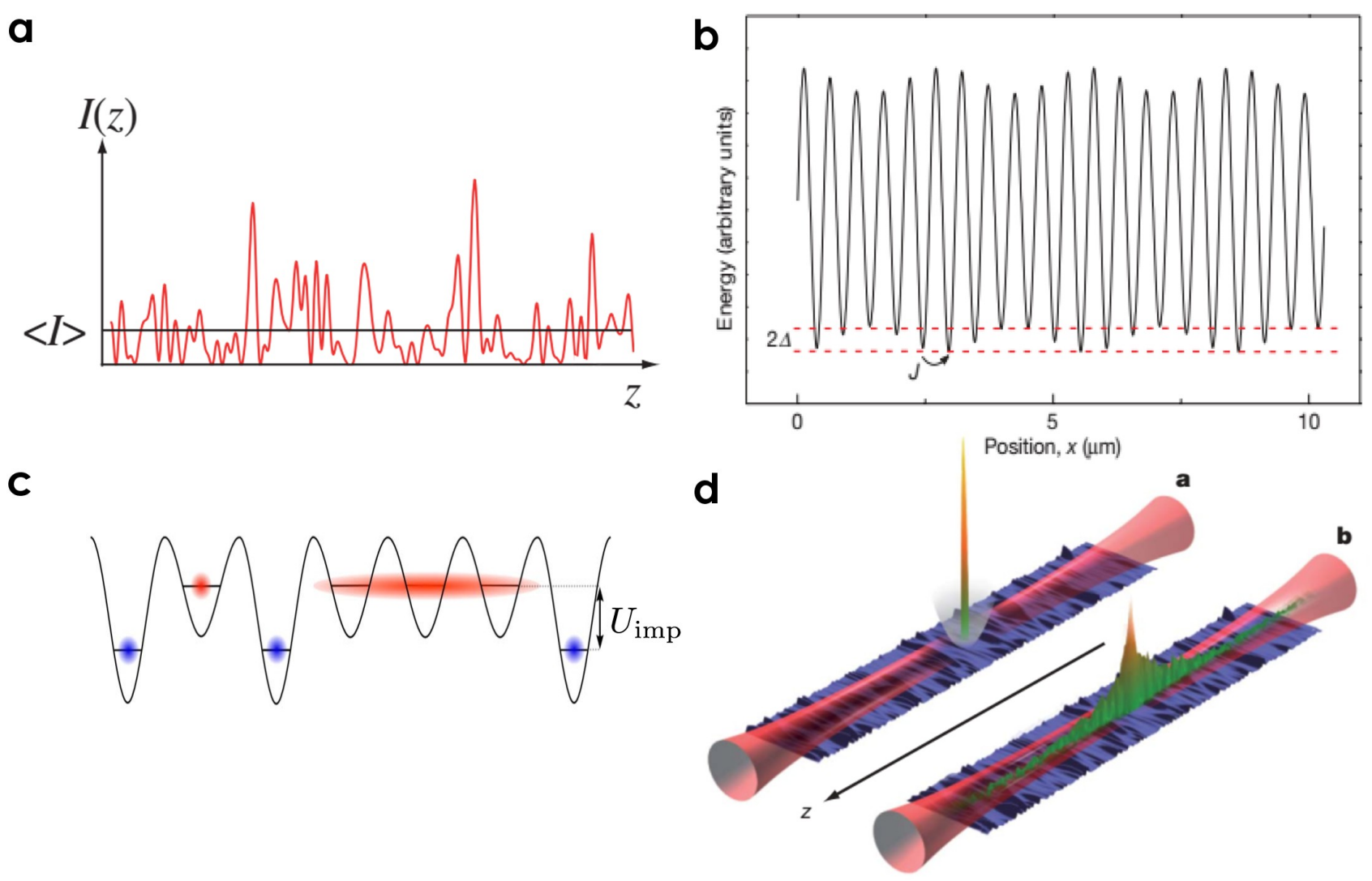

Fig. 4 | **Controlled perturbations through disorder. a** | In the absence of a periodic optical lattice, a random pattern of speckles imprints a random light intensity landscape[239] that randomly fluctuates about a mean value, ⟨*I*⟩, onto the atoms. **b** | In a superposition of a strong optical lattice (in this case, with a lattice constant of 516 nm) and a weaker optical lattice (in this case, with a lattice constant of 431 nm), a quasi-periodic potential is realized. The hopping energy, *J*, varies site-to-site (where *x* is the position) and the maximum shift of the on-site energy is $2\Delta$ (ref.[143]). **c** | A bosonic superfluid in an optical lattice (red) is perturbed owing to localized impurities (blue) that introduce local, effective potential shifts $U_{\mathrm{imp}}$ (ref.[145]). **d** | Anderson localization observed in a speckle pattern experiment. In a 1D trap (red), the speckle pattern (blue) is projected onto a small Bose–Einstein condensate wave packet that is also kept in an additional harmonic confinement (grey) (left). Upon release from the small trap, the cloud expands and eventually localizes owing to the disorder potential (right)[142]. Panel **a** adapted with permission from ref.[239], © IOP Publishing and Deutsche Physikalische Gesellschaft. Reproduced by permission of IOP Publishing. CC BY-NC-SA. Panel **b** adapted from ref.[143], Springer Nature Limited. Panel **c** adapted with permission from ref.[145], APS. Panel **d** adapted from ref.[142], Springer Nature Limited.

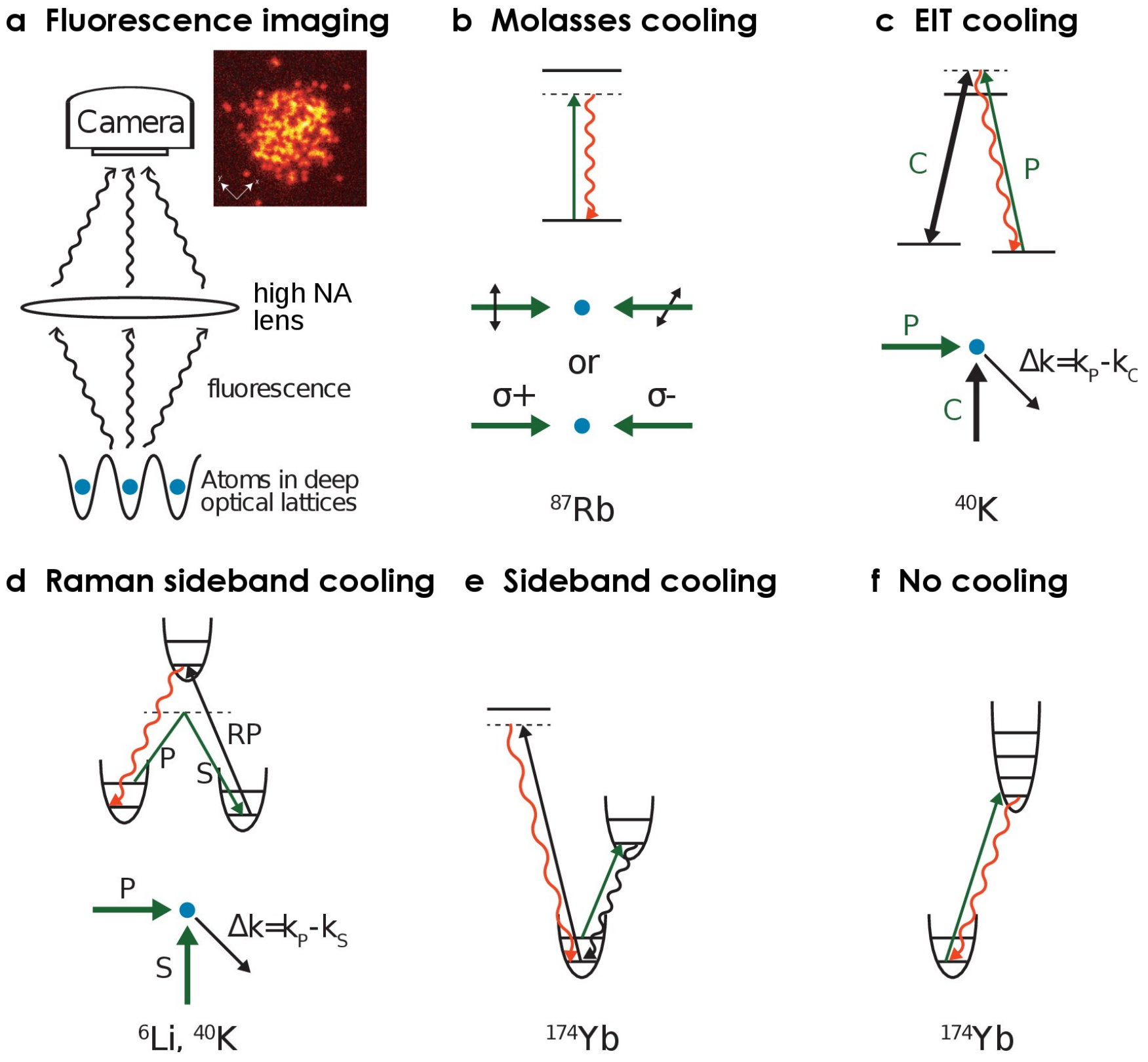

Fig. 5 | **Quantum gas microscope imaging and cooling methods.** In addition to a high-numerical-aperture (NA) lens, cooling schemes are important to obtain a sufficient number of photons. **a** | Fluorescence photons (wiggly arrows) from individual atoms in an optical lattice are first collected by a high-NA lens and then imaged on the camera[2]. The inset shows the raw-data fluorescence image of a weakly interacting Bose–Einstein condensate in an optical lattice. Panels **b–f** show the cooling schemes, restrictions of laser beam configurations (if any) and atomic species for which each cooling method has been demonstrated for achieving site-resolved imaging. **b** | Molasses cooling. The polarizations of the counterpropagating beams are orthogonal (two mutually orthogonal linear polarizations or right- and left-handed $\sigma^+$–$\sigma^-$ polarizations). For alkali-metal atoms, which have magnetic substates in the ground state, polarization gradient cooling is the main cooling mechanism. **c** | Electromagnetically induced transparency (EIT) cooling. A strong coupling beam (C) creates a narrow dressed state, which is driven by the probe beam (P). **d** | Raman sideband cooling. The Raman coupling, consisting of pump (P) and Stokes (S) beams, lowers the vibrational level. The repump beam (RP) prevents the reverse process. An additional repump beam (not shown) is necessary for salvaging atoms from the dark state. For both EIT cooling and Raman sideband cooling, no cooling occurs along the axis perpendicular to the momentum transfer, $\Delta$**k** (where $k_P$, $k_C$ and $k_S$ are the momenta of the pump, coupling and Stokes laser beams, respectively). **e** | Sideband cooling. A narrow optical transition, which exists in alkaline-earth(-like)-metal atoms, such as Yb atoms, makes it possible to resolve the vibrational-level structure and to drive the sideband transition. During detection, a stronger transition can be used in a molasses configuration to obtain a sufficient number of scattered photons while suppressing heating. **f** | No cooling. The lattice confinement for the excited state is so strong that the heating transition is suppressed. Panel **a** (inset) adapted from ref.[2], Springer Nature Limited.

## Box 1 | The optical lattice toolbox

In its most common implementation, an optical lattice is formed by interfering continuous-wave lasers. Most simply, a laser beam with a wavelength $\lambda$ is retroreflected off a mirror, creating a 1D lattice potential, $U(x) = -U_0 \sin^2(2\pi x/\lambda)$ (where $U_0$ is the lattice potential depth, and $x$ is the position of the atoms), that is proportional to the intensity of the laser standing wave. By superimposing 1D lattices in three orthogonal directions, a 3D cubic optical lattice can be created. The periodic potential for the atoms results in the introduction of band structures for the atoms, similar to those of electrons in crystalline materials.

Ultracold atoms trapped in a sufficiently deep lattice potential are described by the Hubbard model (see panel **a** of the figure). For fermionic atoms the Hamiltonian is

$$H_{\text{Fermi-Hubbard}} = -t \sum_{\langle i,j\rangle,\sigma} f^{\dagger}_{i,\sigma} f_{j,\sigma} + U \sum_{i} n^{\text{F}}_{i,\uparrow} n^{\text{F}}_{i,\downarrow} + \sum_{i,\sigma} \epsilon_i n^{\text{F}}_{i,\sigma}$$

where $f^{\dagger}_{i,\sigma}(f_{i,\sigma})$ is the fermionic creation (annihilation) operator for spin $\sigma = \{\uparrow,\downarrow\}$, $n^{F}_{i,\sigma} = f^{\dagger}_{i,\sigma} f_{i,\sigma}$ is the fermionic number operator for $\sigma$-spin at site $i$, $t$ is the hopping matrix element, $U$ is the on-site interaction energy and $\epsilon_i$ is the site-dependent energy offset accounting for weak confinement. $\langle i,j\rangle$ denotes nearest-neighbour sites. Here, it is assumed that the atoms with spin-1/2 occupy a single band of the lattice potential. The Hubbard model features a rich phase diagram, and the competition between kinetic energy and interaction energy leads to quantum phase transitions.

Similar to the case of fermionic atoms, the bosonic counterpart is described by the Bose–Hubbard Hamiltonian,

$$H_{\text{Bose-Hubbard}} = -t \sum_{\langle i,j\rangle,\sigma} b^{\dagger}_{i,\sigma} b_{j,\sigma} + U \sum_{i} n^{\text{B}}_{i}\left(n^{\text{B}}_{i} - 1\right)/2 + \sum_{i,\sigma} \epsilon_i n^{\text{B}}_{i,\sigma}$$

where $b_i(b^{\dagger}_i)$ is the bosonic annihilation (creation) operator and $n^{B}_i = b^{\dagger}_i b_i$ is the number operator for Bosons at site $i$. As the interaction strength $(U/t)$ is increased, the system undergoes a quantum phase transition from the superfluid to the Mott-insulator phase. The Gaussian shape of the laser beams forming the optical lattice leads to an overall harmonic confinement potential, which gives rise to a wedding-cake-like structure of the density distribution in the Mott-insulator phase.

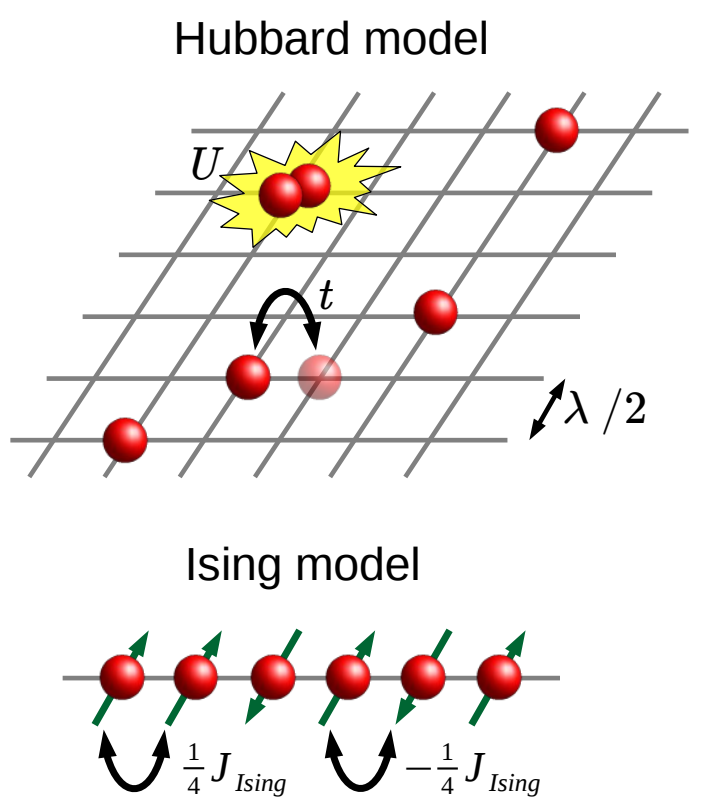

In the limit of half-filling, where one spin-1/2 particle per lattice site is found, and strong interactions $(U/t \gg 1)$, the Fermi–Hubbard model is reduced to the Heisenberg model

$$H_{\text{Heisenberg}} = J \sum_{\langle i,j\rangle} S_i \cdot S_j$$

Here, $S_i = \left(S^x_i, S^y_i, S^z_i\right)$ is the spin operator and $J$ is the nearest-neighbour coupling constant. The coupling is antiferromagnetic for $J > 0$ and ferromagnetic for $J < 0$. The coupling arises from the super-exchange interaction that is given by $J = 4t^2/U$. The Bose–Hubbard model can also be reduced to the anisotropic Heisenberg model[15].

Another important spin model for quantum simulation is the Ising model (see panel **b** of the figure),

$$H_{\text{Ising}} = J_{\text{Ising}} \sum_{\langle i,j\rangle} S^z_j S^z_j + J_{\text{Ising}} \sum_{i} \left(h_x S^x_i - h_z S^z_i\right)$$

where the first term describes the nearest-neighbour interaction that depends only on the $z$-component of the spin, and the second term describes the transverse and longitudinal magnetic field. A Bose–Hubbard model with a tilted potential can be used to emulate the Ising model, wherein the occupation numbers are mapped to spins to observe paramagnetic-to-antiferromagnetic quantum phase transitions[17,18].

# Tools for quantum simulation with ultracold atoms in optical lattices

*Florian Schäfer, Takeshi Fukuhara, Seiji Sugawa, Yosuke Takasu and Yoshiro Takahashi*

## S1. Formation of ultracold atomic gases

Any experiment with ultracold atomic gases for quantum simulation needs to: trap the required atomic species by means of either magnetic or optical fields and cool the trapped atoms to the required ultra-low temperatures. In both steps the effects of light-induced forces are of paramount importance and the necessary theoretical and experimental tools have been developed since the second part of the twentieth century[1–3].

First, for atomic species with very low vapour pressure at room temperature, a 'hot' atomic beam is usually created from a vapour source that, depending on the specific vapour pressure characteristics, is kept at temperatures that range from only slightly above room temperature (for example, Rubidium) to well beyond 1000°C (for example, Erbium or Dysprosium). At this stage the atoms have typical velocities of several hundred meters-per-second. Using the recoil pressure of resonantly counter-propagating laser light in a so-called Zeeman slower the atoms are slowed down to about 10 m/s. At that velocity they can be trapped and laser-cooled in a magneto-optical trap (MOT) to below milli-Kelvin temperatures[4]. A MOT can be directly loaded from the low velocity component of an atomic vapour, if it is large enough, in a glass cell near room temperature. Then the atoms are transferred to either a magnetic or an optical trap where forced evaporative cooling is performed[5]. After the evaporative cooling the atoms are typically below micro-Kelvin temperatures and reach a quantum-degenerate regime, where the atoms undergo a phase transition to a Bose-Einstein condensate (BEC) for bosonic atoms[6,7] or form a Fermi-degenerate gas for fermionic atoms[8]. Ref. [9] used Raman-sideband cooling to achieve BEC, instead of evaporative cooling. Up to now a variety of atomic species have been successfully cooled to quantum-degenerate regimes, including alkali-metal to non-alkali-metal atomic species[10–14], see Table S1a), and mixtures[15–18], Table S1b).

## S2. Measurement of spin-correlations in optical lattices

Without a powerful quantum gas microscope technique, it is usually quite difficult to measure the spin correlation between neighbouring lattice sites. However, the spin correlation can be detected by inducing and observing singlet-triplet oscillations in the double-well structures of an optical superlattice. In this approach, first, the hopping between neighbouring sites is frozen by increasing the optical lattice potential. Then a spin-dependent potential gradient is applied. This induces a coherent oscillation between spin-singlet and -triplet states. Finally, pairs of neighbouring sites are merged into single sites by use of a superlattice[19]. Using for example laser light tuned to an s-wave photoassociation resonance or a magnetic field sweep across an s-wave Feshbach resonance one can selectively remove only those pairs of atoms that are singlet correlated. Note that the atom loss at the beginning of the singlet-triplet oscillation corresponds to the number of initially prepared singlet state pairs and that the loss after half of the oscillation relates to the number of initially prepared triplet state pairs. In this manner spin correlations have been observed successfully[19–21]. Instead of selective removal, singlet-triplet oscillations also have been observed using band-mapping after merging of the neighbouring lattice sites[19,21].

## S3. Technical aspects of quantum gas microscopes

High-resolution imaging for quantum simulation purposes typically requires a resolution to at least match the lattice spacing of the optical lattice that is normally used in such an experiment: several hundreds of nanometres. This is roughly equal to the wavelengths used for imaging, and therefore high-resolution imaging close to the diffraction limit is necessary for site-resolved detection. Although such an imaging is commonly used in biology or chemistry, its realization for ultracold atoms is considerably more challenging, because: the sample of ultracold atoms is prepared in an ultra-high-vacuum environment and it is too fragile to endure heating by illumination of the probe beam. An ultra-high-vacuum environment usually requires a vacuum window with a thickness of more than a millimetre. This is much thicker than a cover glass for a standard microscope, causes serious aberrations especially for high numerical aperture objectives and increases the distance between the microscope lens and the atoms considerably. One solution is to use a custom-made long-working-distance microscope objective with corrections for the specific window thickness[22], see Fig. S1a). Another approach to reduce spatial aberration is to use a 200-micrometer-thickness sapphire window instead of a standard vacuum viewport[23], Fig. S1b). In-vacuum solid immersion lenses are also employed to further enhance the numerical aperture[24], Fig. S1c). Because of the limited working distance of the combined imaging optics the atoms need to be prepared close to the vacuum window or the solid immersion lens. To this end, atoms are transported from the laser cooling region to an imaging region by either magnetic forces[25] or by optical potentials such as optical tweezers[26] or moving optical lattices[27].

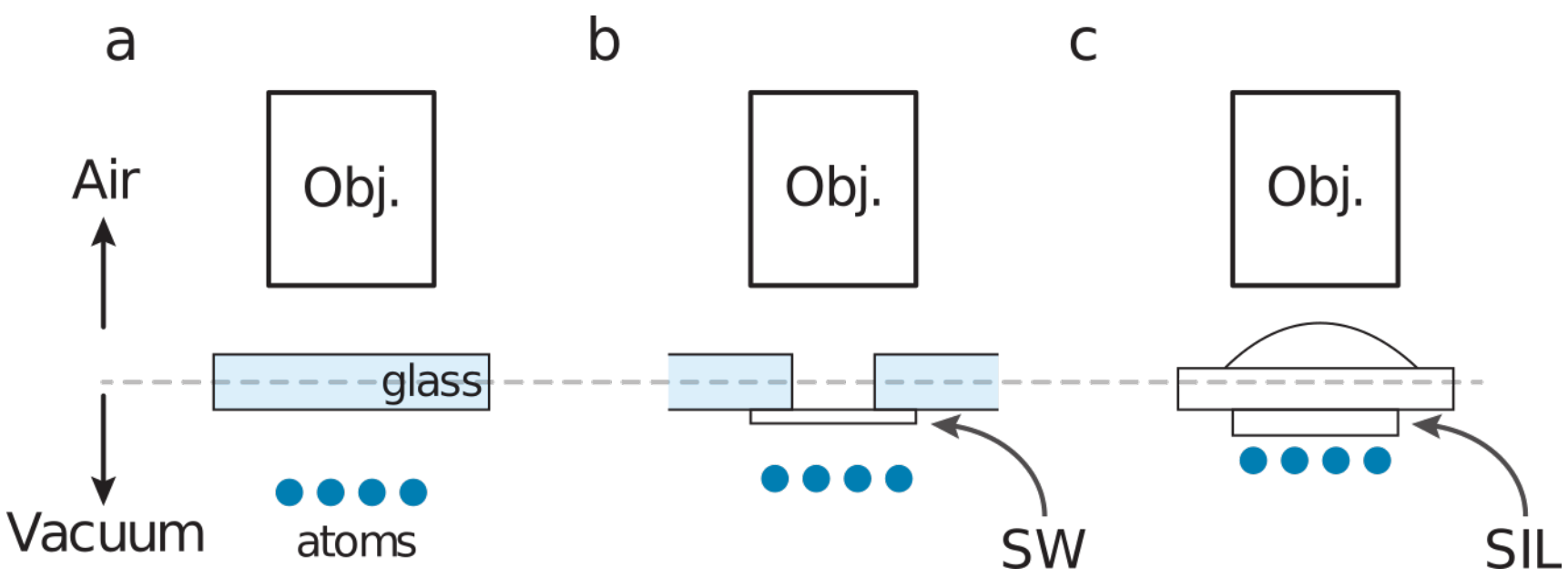

Fig. S1 | **Variants of quantum gas microscopes** . Some possibilities to achieve high-resolution imaging even though an air-vacuum transition has to be included into the imaging setup **a** | Custom-made objective including corrections for the thick vacuum window glass. **b** | Objective with 200-micrometer-thickness sapphire window (SW). **c** | Objective with additional solid immersion lens (SIL).

## S4. Comparison of synthetic gauge fields and spin-orbit coupling implementations

We compare different systems that have experimentally implemented synthetic gauge fields in real space and spin-orbit coupling. Particularly, we focus on three major methods for implementing synthetic gauge fields (Table S2a) and 1D and 2D spin-orbit coupling, including non-lattice systems (Table S2b). For the synthetic dimensional systems a comparison list can be found in another review[28].

Table S1 | Atomic species and mixtures thereof that have been brought to quantum degeneracy.

a) Atomic species for which Bose-Einstein condensation or Fermi degeneracy has been achieved.

| Atomic species | Boson or fermion | Publication year | Reference |
|---|---|---|---|
| $^{1}$H | boson | 1998 | Ref. [29] |
| $^{3}$He* | fermion | 2006 | Ref. [30] |
| $^{4}$He* | boson | 2001 | Refs [31,32] |
| $^{6}$Li | fermion | 2001 | Ref. [33] |
| $^{7}$Li | boson | 1995 | Ref. [34] |
| $^{23}$Na | boson | 1995 | Ref. [35] |
| $^{39}$K | boson | 2007 | Ref. [36] |
| $^{40}$K | fermion | 1999 | Ref. [8] |
| $^{41}$K | boson | 2001 | Ref. [37] |
| $^{40}$Ca | boson | 2009 | Ref. [38] |
| $^{52}$Cr | boson | 2005 | Ref. [11] |
| $^{53}$Cr | fermion | 2015 | Ref. [39] |
| $^{85}$Rb | boson | 2000 | Ref. [40] |
| $^{87}$Rb | boson | 1995 | Ref. [6] |
| $^{84}$Sr | boson | 2009 | Refs [12,41] |
| $^{86}$Sr | boson | 2010 | Ref. [42] |
| $^{87}$Sr | fermion | 2010 | Refs [43,44] |
| $^{88}$Sr | boson | 2010 | Ref. [45] |
| $^{133}$Cs | boson | 2003 | Ref. [46] |
| $^{161}$Dy | fermion | 2012 | Ref. [47] |
| $^{164}$Dy | boson | 2011 | Ref. [13] |
| $^{166}$Er | boson | 2016 | Ref. [48] |
| $^{168}$Er | boson | 2012 | Ref. [14] |
| $^{167}$Er | fermion | 2014 | Ref. [49] |
| $^{168}$Yb | boson | 2011 | Ref. [50] |
| $^{170}$Yb | boson | 2007 | Ref. [51] |
| $^{171}$Yb | fermion | 2010 | Ref. [17] |
| $^{173}$Yb | fermion | 2007 | Ref. [52] |
| $^{174}$Yb | boson | 2003 | Ref. [10] |
| $^{176}$Yb | boson | 2009 | Ref. [16] |

b) Combination of atomic species for which quantum degenerate mixtures have been achieved (not including quantum-degenerate molecules). †The letter B (F) after the atomic symbols indicates that the isotope is a boson (fermion).

| Atomic species† | Publication year | Reference |
|---|---|---|
| $^{3}$He* (F) + $^{4}$He* (B) | 2006 | Ref. [30] |
| $^{6}$Li (F) + $^{7}$Li (B) | 2001 | Ref. [33] |
| $^{6}$Li (F) + $^{23}$Na (B) | 2002 | Ref. [15] |
| $^{6}$Li (F) + $^{40}$K (F) + $^{41}$K (B) | 2011 | Ref. [53] |
| $^{6}$Li (F) + $^{40}$K (F) + $^{87}$Rb (B) | 2008 | Ref. [54] |
| $^{6}$Li (F) + $^{87}$Rb (B) | 2005 | Ref. [55] |
| $^{6}$Li (F) + $^{133}$Cs (B) | 2017 | Ref. [56] |
| $^{6,7}$Li (F, B) + $^{174,173}$Yb (B, F) | 2011, 2018 | Refs [18,57], Ref. [58] |
| $^{23}$Na (B) + $^{39}$K (B) | 2018 | Ref. [59] |
| $^{23}$Na (B) + $^{40}$K (F) | 2012 | Ref. [60] |
| $^{23}$Na (B) + $^{87}$Rb (B) | 2016 | Ref. [61] |
| $^{39}$K (B) + $^{87}$Rb (B) | 2015 | Ref. [62] |
| $^{40}$K (F) + $^{87}$Rb (B) | 2002 | Ref. [63] |
| $^{41}$K (B) + $^{87}$Rb (B) | 2002 | Ref. [64] |
| $^{40}$K (F) + $^{161}$Dy (F) | 2018 | Ref. [65] |
| $^{85}$Rb (B) + $^{87}$Rb (B) | 2008 | Ref. [66] |
| $^{87}$Rb (B) + $^{133}$Cs (B) | 2011 | Refs [67,68] |
| $^{87}$Rb (B), $^{84}$Sr (B), $^{88}$Sr (B) | 2013 | Ref. [69] |
| $^{84}$Sr (B), $^{86}$Sr (B), $^{87}$Sr (F), $^{88}$Sr (B) | 2010, 2013 | Ref. [44], Ref. [70] |
| $^{166}$Er (B), $^{168}$Er (B), $^{167}$Er (F) + $^{161}$Dy (F), $^{164}$Dy (B) | 2018 | Ref. [71] |
| $^{171}$Yb (F) + $^{87}$Rb (B) | 2015 | Ref. [72] |
| $^{168}$Yb (B), $^{170}$Yb (B), $^{171}$Yb(F), $^{173}$Yb (F), $^{174}$Yb (B), $^{176}$Yb (B) | 2010, 2007, 2011 | Ref. [17], Ref. [73], Ref. [50] |

Table S2 | Implementations of synthetic gauge fields and spin-orbit couplings (SOC) in real space.

a) Implementation of synthetic gauge fields in real space.

| **Method** | **System** | **Features** |
|---|---|---|
| 1. Raman-coupled hyperfine states | optical trap | effective magnetic field for BEC[74] |
| | rf+Raman dressed 1D lattice | Peierls substitution in 1D lattice[75] |
| 2. Raman-induced tunneling | superlattice | staggered ‘magnetic’ flux[76] |
| | 2D tilted optical lattice | uniform effective magnetic field, Harper-Hofstadter model[77–80] |
| 3. lattice shaking | 1D optical lattice | Peierls substitution in 1D lattice[81] |
| | triangular lattice | staggered effective magnetic flux, classical spin model[82] |
| | honeycomb lattice | Haldane model[83,84] |

b) Implementation of one-dimensional SOC.

| Atoms | Coupled states | Scheme and features |
|---|---|---|
| $^{87}$Rb (B) | hyperfine spin states | Raman coupling[85] |
| $^{40}$K (F) | hyperfine spin states | Raman coupling[86] |
| $^{6}$Li (F) | hyperfine spin states | Raman coupling[87], 1D SOC lattice by rf+Raman dressing |
| $^{161}$Dy (F) | electronic spin states | large electronic spin[88], large dipolar interaction |
| $^{87}$Sr (F) | ground/excited electronic states | clock transition[89], magic wavelength optical lattice |
| $^{173}$Yb (F) | ground/excited electronic states | clock transition[90], magic-wavelength optical lattice |
| $^{171}$Yb (F) | nuclear spin states | ultra-narrow transition[91] |
| $^{87}$Rb (B) | hyperfine spin states | Raman coupling, dynamical SOC in an optical cavity[92] |
| $^{173}$Yb (F) | nuclear spin states | Raman coupling[93], 1D optical Raman lattice[94] |
| $^{87}$Rb (B) | sublattices of optical superlattice | Raman laser induced tunneling[95] |

c) Implementation of two-dimensional SOC.

| Atoms | Coupled states | Scheme and features |
|---|---|---|
| $^{87}$Rb (B) | hyperfine spin states | 2D optical Raman lattice[96,97], topological lattice band |
| $^{40}$K (F) | hyperfine spin states | Raman coupling[98,99] |
| $^{173}$Yb (F) | nuclear spin states | 2D optical Raman lattice, additional coupling, nodal-line semi-metal phase[100] |
| $^{87}$Rb (B) | synthetic clock states | Raman coupling[101] |

## Supplementary Information References